\documentclass[10pt,twocolumn]{article} 
\usepackage{simpleConference}
\usepackage{listings}
\usepackage{amsmath}
\usepackage{amsfonts}
\usepackage{tikz}
\usepackage[toc,page]{appendix}
\usepackage{mathrsfs}
\usepackage{paralist}
\usepackage[shortlabels]{enumitem}
\usepackage{lipsum, adjustbox}
\usepackage{multirow}
\usepackage{multicol}
\usepackage{tablefootnote}
\usepackage{graphicx}
\usepackage{capt-of}
\usepackage{pgf-umlsd}
\usepackage{breqn}
\usepackage{graphicx}
\usepackage{paralist}
\newtheorem{mydef, definition}{Definition}

\newtheorem{theorem,lemma, proposition, remark}{Theorem,Lemma, Proposition, Remark}
\usepackage[ruled,vlined]{algorithm2e}
\include{pythonlisting}
\usepackage{color,soul}
\usepackage{hyperref}

\usepackage{cleveref}
\usepackage{algpseudocode}
\usepackage{caption}
\usepackage{url}
\usepackage{enumitem}
\setlist{leftmargin=5pt,labelindent=5pt}
\setlist[enumerate]{wide=0pt, leftmargin=15pt, labelwidth=15pt, align=left}
\usepackage{balance}

\usepackage{lipsum, adjustbox}

\usetikzlibrary{shapes.geometric}

\colorlet{linecol}{black!75}

\hypersetup{
	hypertexnames=true, linkcolor=blue, anchorcolor=black,
	citecolor=blue, urlcolor=blue  
}

\makeatletter
\newsavebox\myboxA
\newsavebox\myboxB
\newlength\mylenA
\newcommand*\xoverline[2][1.00]{%
	\sbox{\myboxA}{$\m@th#2$}%
	\setbox\myboxB\null
	\ht\myboxB=\ht\myboxA%
	\dp\myboxB=\dp\myboxA%
	\wd\myboxB=#1\wd\myboxA
	\sbox\myboxB{$\m@th\overline{\copy\myboxB}$}
	\setlength\mylenA{\the\wd\myboxA}
	\addtolength\mylenA{-\the\wd\myboxB}%
	\ifdim\wd\myboxB<\wd\myboxA%
	\rlap{\hskip 0.5\mylenA\usebox\myboxB}{\usebox\myboxA}%
	\else
	\hskip -0.5\mylenA\rlap{\usebox\myboxA}{\hskip 0.5\mylenA\usebox\myboxB}%
	\fi}
\makeatother
\newcommand{\overbar}[1]{\mkern 1.5mu\overline{\mkern-1.5mu#1\mkern-1.5mu}\mkern 1.5mu}
\begin{document}

\title{Split Ways: Privacy-Preserving Training of Encrypted Data Using Split Learning}

\author{Tanveer Khan, Khoa Nguyen, Antonis Michalas 
}

\maketitle
\thispagestyle{empty}

\begin{abstract}
  Split Learning (SL) is a new collaborative learning technique that allows participants, e.g. a client and a server, to train machine learning models without the client sharing raw data. In this setting, the client initially applies its part of the machine learning model on the raw data to generate activation maps and then sends them to the server to continue the training process. Previous works in the field demonstrated that reconstructing activation maps could result in privacy leakage of client data. In addition to that, existing mitigation techniques that overcome the privacy leakage of SL prove to be significantly worse in terms of accuracy. In this paper, we improve upon previous works by constructing a protocol based on U-shaped SL that can operate on homomorphically encrypted data. More precisely, in our approach, the client applies Homomorphic Encryption (HE) on the activation maps before sending them to the server, thus protecting user privacy. This is an important improvement that reduces privacy leakage in comparison to other SL-based works. Finally, our results show that, with the optimum set of parameters, training with HE data in the U-shaped SL setting only reduces accuracy by 2.65\% compared to training on plaintext. In addition, raw training data privacy is preserved.
\end{abstract}

\section{Introduction}
\label{sec:intro}

Machine Learning (ML) models have attracted global adulation and are used in a plethora of applications such as medical diagnosis, pattern recognition, and credit risk assessment. However, applications and services 
using ML are often breaching user privacy. As a result, the need to preserve the confidentiality and privacy of individuals and maintain user trust has gained extra attention. This is not only because of the technological advancements that privacy-preserving machine learning (PPML) can offer, but also due to its potential societal impact (i.e.\ 
building fairer, democratic and unbiased societies).

Split Learning (SL) and Federated Learning (FL) are the two methods of collaboratively training a model derived from distributed data sources without sharing raw data~\cite{vepakomma2019reducing}. In FL, every client runs a copy of the entire model on its data. The server receives updated weights from each client and aggregates them. 
The SL model divides the neural network into two parts: the client-side and the server-side~\cite{singh2019detailed}. 
SL is used for training Deep Neural Networks (DNN) among multiple data sources, while mitigating the need to directly share raw labeled data with collaboration parties. 
The advantages of SL are multifold: \begin{inparaenum}[\it (i)] \item it allows users to train ML models without sharing their raw data with a server running part of a DNN model, thus preserving
	user privacy. \item it protects both the client and the server from revealing their parts of the model, and \item it reduces the client's computational overhead by not running the entire model (i.e. utilizing a smaller number of layers)~\cite{vepakomma2018split}\end{inparaenum}. 
Though SL offers an extra layer of privacy protection by definition, there 
are no works 
exploring how it is combined with popular techniques that promise to preserve user privacy (e.g.\ encryption). In~\cite{abuadbba2020can}, the authors studied 
whether SL can 
handle 
sensitive time-series data and demonstrated that SL alone is \textit{insufficient} 
when performing privacy-preserving training for 1-dimensional (1D) CNN models. More precisely, the authors showed 
raw data can be reconstructed from the activation maps of the intermediate split layer. The authors also employed two mitigation techniques, adding 
hidden layers and applying differential privacy to reduce 
privacy leakage. However, based on the results, none of these techniques can effectively reduce 
privacy leakage from all channels of the SL activation. Furthermore, both these techniques 
result in 
reducing the joint model's accuracy. 

In this work, we construct a model that uses Homomorphic Encryption (HE)~\cite{cheon2017homomorphic} to mitigate 
privacy leakage in SL. In our proposed model, the client first encrypts the activation maps and then sends the encrypted activation maps to the server. The encrypted activation maps do \textit{not} reveal anything about the raw data (i.e.\ it is \textit{not} possible to reconstruct the original raw data from the encrypted activation maps).


\smallskip
\noindent\textbf{Vision} 
\label{par:Vision}
AI systems have proven surpass people in recognizing abnormalities such as tumours on X-rays and ultrasound scans~\cite{wooldridge2020road}. In addition to that, machines can reliably make diagnoses equal to those of human experts. All the evidence indicates that we can now build systems that achieve human expert performance in analyzing medical data -- systems allowing humans to send their medical data to a remote AI service and receive an accurate automated diagnosis. An intelligent and efficient AI healthcare system of this type offers a great potential since it can improve the health of humans but also have an important social impact. However, these opportunities come with certain pitfalls, mainly concerning privacy. With this in mind, we have designed a system that analyzes images in a privacy-preserving way. More precisely, we show how encrypted images can be analyzed with high accuracy without leaking information about their actual content. While this is still far from our big dream (namely automated AI diagnosis) we still believe it is an important step that will eventually pave the way towards our timate goal.

\smallskip
\noindent \textbf{Contributions} 
\label{par:ourcontribution}
The main contributions of this paper are the following:  
\begin{itemize}[noitemsep,topsep=2pt]
	
	\item We designed a simplified version of the 1D CNN model presented in~\cite{abuadbba2020can} and we are using it to classify the ECG signals~\cite{moody2001impact} in both local and split learning settings.  More specifically, we construct a U-shaped split 1D CNN model and experiment using 
	plaintext activation maps sent from the client to the server. 
	Through the U-shaped 1D CNN model, 
	clients do \textit{not} need to share the input training samples and the ground truth labels with the server -- this is an important improvement that reduces privacy leakage compared to~\cite{abuadbba2020can}.
	
	\item We constructed the HE version of the U-shaped SL technique. In the encrypted U-shaped SL model, the client encrypts the activation map using HE and sends it to the server. The core advantage of the HE encrypted U-shaped SL over the plaintext U-shaped SL is that the server performs computation over the encrypted activation maps.
	\item To assess the applicability of our framework, we performed experiments on 
	a heartbeat datasets: the MIT-DB~\cite{moody2001impact} 
	For this dataset, we experimented with activation maps of~256 
	for both plaintext and homomorphically encrypted activation maps  
	and we have measured the model's performance by considering 
	training
	duration, 
	test accuracy, and 
	communication cost. 
	
\end{itemize}

\section{Related Work}
\label{sec:relatedwork}


The SL approach proposed by Gupta and Raskar~\cite{gupta2018distributed} offers 
a number of significant advantages over FL. Similar to FL~\cite{yang2019federated}, SL does \textit{not} share raw data. In addition
, it has the benefit of \textit{not} disclosing the model's architecture and weights. For example, 
~\cite{gupta2018distributed} predicted that 
reconstructing raw data on the client-side , while using 
SL 
would be difficult. In addition, the authors of 
~\cite{vepakomma2018split}
employed the SL model to the healthcare applications to protect the users' personal data. Vepakomma \textit{et al.} found that SL outperforms FL in terms of accuracy~\cite{vepakomma2018split}.

Initially, it was believed that SL is a promising approach in terms of client raw data protection, 
however, SL provides data privacy 
on the grounds that only intermediate activation maps are shared between the parties. Different studies showed the possibility of privacy leakage in SL. 
In~\cite{vepakomma2019reducing}, the authors analyzed the privacy leakage of SL and found a considerable leakage from the split layer in the 2D~CNN model. Furthermore, the authors mentioned that it is possible to reduce the distance correlation (a measure of dependence) between the split layer and raw data by slightly scaling the weights of all layers before the split. This type of scaling works well in models 
with a large number of hidden layers before the split. 

The work of Abuadbba \textit{et al.}~\cite{abuadbba2020can} is the first study 
exploring 
whether SL can deal with time-series data. It is dedicated to investigating \begin{inparaenum}[\it (i)] \item 
	whether an SL can achieve the same model accuracy for a 1D CNN model compared to the non-split version and 
	\item 
	whether it can be used to protect privacy in sequential data. \end{inparaenum} According to the results, SL can be applied to a model without the model classification accuracy degradation. As for 
the second question, the authors proved 
it is possible to reconstruct the raw data (personal ECG signal) in the 1D CNN model using SL by proposing a privacy assessment framework. 
They suggested three metrics: visual invertibility, distance correlation, and dynamic time warping. 
The results showed that 
when SL is directly adopted into 1D CNN models for time series data could result in significant privacy leakage. Two mitigation techniques were employed to limit the potential privacy leakage in SL: \begin{inparaenum}[\it (i)] \item increasing the number of layers before the split on the client-side and \item applying differential privacy to the split layer activation before sending the activation map to the server\end{inparaenum}. However, both 
techniques suffer from a loss of model accuracy, particularly when differential privacy is used. The strongest differential privacy can increase the dissimilarity between the activation map and the corresponding raw data. However, \textit{it degrades the classification accuracy significantly from 98.9\% to 50\%.}

In~\cite{abuadbba2020can}, during the forward propagation, the client sends the activation map in plaintext to the server, where 
the server can easily reconstruct the original raw data from the activated vector of the split layer leading to clear privacy leakage. In our work, we constructed a training protocol, where, instead of sending 
plaintext activation maps, the client first conducts an encryption using HE and then sends said maps 
to the server. In this way, the server is unable to reconstruct the original raw data, but can still 
perform a computation on the encrypted activation maps and 
realize the training process.

\section{Architecture}
\label{sec:architecture}
In this section,
we first describe the non-split version or local model of the 1D CNN used to classify the ECG signal. Then, we discuss the process of splitting this local model into a U-shaped split model. Furthermore, we also describe the involved parties (a client and a server) in the training process of the split model, focusing on their roles and the parameters 
assigned to them throughout the training process.

\subsection{1D CNN Local Model Architecture}
\label{subsec:local_model}

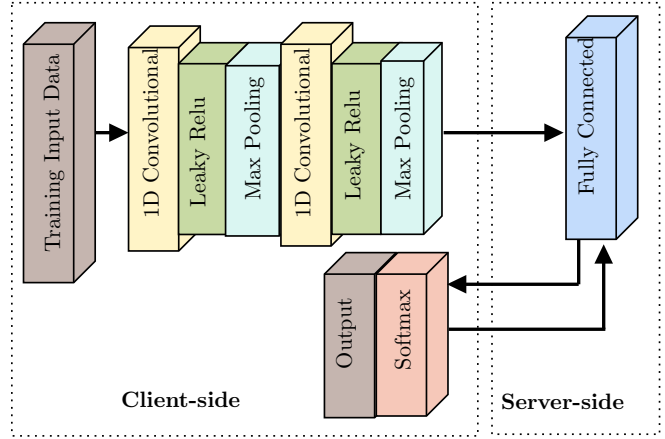
\begin{figure}
	\begin{adjustbox}{width=0.5\textwidth}
		\tikzset{every picture/.style={line width=0.75pt}} 
		\begin{tikzpicture}[x=0.75pt,y=0.75pt,yscale=-1,xscale=1]
			\draw  [fill={rgb, 255:red, 255; green, 244; blue, 199 }  ,fill opacity=1 ] (167,155.9) -- (179.9,143) -- (210,143) -- (210,265.1) -- (197.1,278) -- (167,278) -- cycle ; \draw   (210,143) -- (197.1,155.9) -- (167,155.9) ; \draw   (197.1,155.9) -- (197.1,278) ;
			\draw  [fill={rgb, 255:red, 200; green, 218; blue, 164 }  ,fill opacity=1 ] (197,165.6) -- (209.6,153) -- (239,153) -- (239,255.5) -- (226.4,268.1) -- (197,268.1) -- cycle ; \draw   (239,153) -- (226.4,165.6) -- (197,165.6) ; \draw   (226.4,165.6) -- (226.4,268.1) ;
			\draw  [fill={rgb, 255:red, 218; green, 246; blue, 242 }  ,fill opacity=1 ] (225.2,166.8) -- (239,153) -- (271.2,153) -- (271.2,255.2) -- (257.4,269) -- (225.2,269) -- cycle ; \draw   (271.2,153) -- (257.4,166.8) -- (225.2,166.8) ; \draw   (257.4,166.8) -- (257.4,269) ;
			\draw  [fill={rgb, 255:red, 197; green, 181; blue, 175 }  ,fill opacity=1 ] (104,153.9) -- (116.9,141) -- (147,141) -- (147,284.1) -- (134.1,297) -- (104,297) -- cycle ; \draw   (147,141) -- (134.1,153.9) -- (104,153.9) ; \draw   (134.1,153.9) -- (134.1,297) ;
			\draw  [fill={rgb, 255:red, 255; green, 244; blue, 199 }  ,fill opacity=1 ] (258.4,153.9) -- (271.3,141) -- (301.4,141) -- (301.4,262.1) -- (288.5,275) -- (258.4,275) -- cycle ; \draw   (301.4,141) -- (288.5,153.9) -- (258.4,153.9) ; \draw   (288.5,153.9) -- (288.5,275) ;
			\draw  [fill={rgb, 255:red, 200; green, 218; blue, 164 }  ,fill opacity=1 ] (289,165.6) -- (301.6,153) -- (331,153) -- (331,255.4) -- (318.4,268) -- (289,268) -- cycle ; \draw   (331,153) -- (318.4,165.6) -- (289,165.6) ; \draw   (318.4,165.6) -- (318.4,268) ;
			\draw  [fill={rgb, 255:red, 218; green, 246; blue, 242 }  ,fill opacity=1 ] (318.4,163.98) -- (329.38,153) -- (355,153) -- (355,257.02) -- (344.02,268) -- (318.4,268) -- cycle ; \draw   (355,153) -- (344.02,163.98) -- (318.4,163.98) ; \draw   (344.02,163.98) -- (344.02,268) ;
			\draw  [fill={rgb, 255:red, 195; green, 220; blue, 252 }  ,fill opacity=1 ] (429.4,149.9) -- (442.3,137) -- (472.4,137) -- (472.4,258.1) -- (459.5,271) -- (429.4,271) -- cycle ; \draw   (472.4,137) -- (459.5,149.9) -- (429.4,149.9) ; \draw   (459.5,149.9) -- (459.5,271) ;
			\draw  [fill={rgb, 255:red, 197; green, 181; blue, 175 }  ,fill opacity=1 ] (284.5,291.9) -- (297.4,279) -- (327.5,279) -- (327.5,365.16) -- (314.6,378.06) -- (284.5,378.06) -- cycle ; \draw   (327.5,279) -- (314.6,291.9) -- (284.5,291.9) ; \draw   (314.6,291.9) -- (314.6,378.06) ;
			\draw  [fill={rgb, 255:red, 246; green, 200; blue, 185 }  ,fill opacity=1 ] (315.6,291.9) -- (328.5,279) -- (358.6,279) -- (358.6,364.17) -- (345.7,377.07) -- (315.6,377.07) -- cycle ; \draw   (358.6,279) -- (345.7,291.9) -- (315.6,291.9) ; \draw   (345.7,291.9) -- (345.7,377.07) ;
			\draw [line width=1.5]    (147,207) -- (163,207) ;
			\draw [shift={(167,207)}, rotate = 180] [fill={rgb, 255:red, 0; green, 0; blue, 0 }  ][line width=0.08]  [draw opacity=0] (11.61,-5.58) -- (0,0) -- (11.61,5.58) -- cycle    ;
			\draw [line width=1.5]    (356,207) -- (424,207) ;
			\draw [shift={(428,207)}, rotate = 180] [fill={rgb, 255:red, 0; green, 0; blue, 0 }  ][line width=0.08]  [draw opacity=0] (11.61,-5.58) -- (0,0) -- (11.61,5.58) -- cycle    ;
			\draw [line width=1.5]    (437,298) -- (362,298) ;
			\draw [shift={(358,298)}, rotate = 360] [fill={rgb, 255:red, 0; green, 0; blue, 0 }  ][line width=0.08]  [draw opacity=0] (11.61,-5.58) -- (0,0) -- (11.61,5.58) -- cycle    ;
			\draw  [dash pattern={on 0.84pt off 2.51pt}] (97,129) -- (376.5,129) -- (376.5,389) -- (97,389) -- cycle ;
			\draw  [dash pattern={on 0.84pt off 2.51pt}] (385,129) -- (485.5,129) -- (485.5,388) -- (385,388) -- cycle ;
			\draw [line width=1.5]    (437,271) -- (437,298) ;
			\draw [line width=1.5]    (359,325) -- (453,325) ;
			\draw [line width=1.5]    (452,325) -- (452,278) ;
			\draw [shift={(452,274)}, rotate = 90] [fill={rgb, 255:red, 0; green, 0; blue, 0 }  ][line width=0.08]  [draw opacity=0] (11.61,-5.58) -- (0,0) -- (11.61,5.58) -- cycle    ;
			\draw (174.42,260.07) node [anchor=north west][inner sep=0.75pt]  [rotate=-269.64] [align=left] {1D Convolutional};
			\draw (265.42,260.07) node [anchor=north west][inner sep=0.75pt]  [rotate=-269.64] [align=left] {1D Convolutional};
			\draw (202.42,253.07) node [anchor=north west][inner sep=0.75pt]  [rotate=-269.64] [align=left] {Leaky Relu};
			\draw (295.42,249.07) node [anchor=north west][inner sep=0.75pt]  [rotate=-269.64] [align=left] {Leaky Relu};
			\draw (235.42,249.07) node [anchor=north west][inner sep=0.75pt]  [rotate=-269.64] [align=left] {Max Pooling};
			\draw (325.42,249.07) node [anchor=north west][inner sep=0.75pt]  [rotate=-269.64] [align=left] {Max Pooling};
			\draw (161,361) node [anchor=north west][inner sep=0.75pt]   [align=left] {\textbf{Client-side}};
			\draw (389,362) node [anchor=north west][inner sep=0.75pt]   [align=left] {\textbf{Server-side}};
			\draw (325.42,350.83) node [anchor=north west][inner sep=0.75pt]  [rotate=-269.64] [align=left] {Softmax};
			\draw (290.42,350.28) node [anchor=north west][inner sep=0.75pt]  [rotate=-269.64] [align=left] {Output};
			\draw (115.42,280.07) node [anchor=north west][inner sep=0.75pt]  [rotate=-269.64] [align=left] {Training Input Data};
			\draw (435.42,246.07) node [anchor=north west][inner sep=0.75pt]  [rotate=-269.64] [align=left] {Fully Connected};
			
		\end{tikzpicture}
	\end{adjustbox}
	\caption{U-shaped Split-Learning}
	\label{fig:u-shapedSL}
\end{figure}

We first implement and successfully reproduce the local model results~\cite{abuadbba2020can}. This model contains two Conv1D layers and two FC layers. The 
optimal test accuracy that this model achieves is 98.9\%. We implement a simplified version where the model has one less FC layer compared to the model from~\cite{abuadbba2020can}. Our local model consists of all the layer of~\autoref{fig:u-shapedSL} 
without any split between the client and the server. As can be seen in \autoref{fig:u-shapedSL}, we limit our model to two Conv1D layers and 
one linear layer as we aim 
to reduce 
computational costs when 
HE is applied on activation maps in the model's split version. 
Reducing the number of FC layers leads to a drop in the accuracy of the model. The best test accuracy we obtained after training our local model for 10 epochs with a batch size of 4 is 92.84\%. \textit{Although reducing the number of layers affects the model's accuracy, 
	it is not within our goals to demonstrate how successful our ML model is for this task; instead, our focus is to construct a split model where training and evaluation on encrypted data are comparable to training and evaluation on plaintext data.}


In \autoref{sec:performance}, we 
detail the results for the non-split version 
and compare them with the split version.

\subsection{U-shaped Split 1D CNN Model}
\label{subsec:u_shaped_split}


The split learning protocol consists of two parties: the client and server. We split the local 1D CNN into multiple parts, where each party trains its part(s) and communicates with 
others to complete the overall training procedure. More specifically, we construct the U-shaped split 1D CNN in such a way that 
the first few 
as well as the last layer are on the client-side, while the remaining layers are on the server-side. 

\paragraph*{\textbf{Actors in the Split Learning Model}}
\label{subsec:slActors}
As mentioned earlier, in our split learning setting, we have two involved parties: the client and the server. Each party plays a specific role and has access to certain parameters. More specifically, their roles and accesses are described as following
\begin{itemize}[noitemsep,topsep=2pt]
	\item Client: In the plaintext version, the client holds two Conv1D layers and 
	can access 
	their weights and biases in plaintext. Other layers (Max Pooling layers, Leaky ReLU layers, Softmax layer) do not have weights and biases. Apart from these, in the HE encrypted version, the client is also responsible for generating the context for HE and has access to all 
	context parameters (Polynomial modulus ($\mathcal{P}$), Coefficient modulus ($\mathcal{C}$), Scaling factor ($\Delta$), Public key ($\mathsf{pk}$) and Secret key ($\mathsf{sk}$)). Note that for both training on plaintext and encrypted activation maps, the raw data examples $\mathbf{x}$'s and their corresponding labels $\mathbf{y}$'s reside on the client side and are never sent to the server during the training process.
	\item Server: In our model, the computation performed on the server-side is limited to only one linear layer. Hence, the server 
	can exclusively access 
	the weights and biases of this linear layer. Regarding the HE context parameters, the server has access to $\mathcal{P}$, $\mathcal{C}$, $\Delta$, and $\mathsf{pk}$ shared by the client, with the exception of 
	the $\mathsf{sk}$. 
	Not holding the $\mathsf{sk}$, the server cannot decrypt the HE encrypted activation maps sent from the client. The hyperparameters shared between the client and the server are the learning rate ($\eta$), batch size ($n$), number of batches to be trained ($N$), and number of training epochs ($E$).
\end{itemize}

\section{Split Model Training Protocols}
\label{sec:slProtocol}
In this section, we first present the protocol for training the U-shaped split 1D CNN on plaintext activation maps, followed by the protocol for training the U-shaped split 1D CNN on encrypted activation maps.

\subsection{Training U-shaped Split Learning with Plaintext Activation Maps}
\label{subsection: unencryptedactivation}

We have used ~\autoref{alg:client} and~\autoref{alg:server} to train the U-shaped split 1D CNN reported in \autoref{subsec:u_shaped_split}. First, the client and server start the socket initialization process and synchronize the hyperparameters $\eta, n, N, E$. They also initialize the weights ($\boldsymbol{w}^{i}$) and biases  ($\boldsymbol{b}^{i}$) of their layers according to $\Phi$.

During the forward propagation phase, the client forward-propagates the input $\mathbf{x}$ until the $l^{th}$ layer and sends the activation $\mathbf{a}^{(l)}$ to the server. The server continues 
to forward propagate and sends the output $\mathbf{a}^{(L)}$ to the client. Next, the client applies the Softmax function on $\mathbf{a}^{(L)}$ to get $\mathbf{\hat{y}}$ and calculates the error $J = \mathcal{L}(\mathbf{\hat{y}}, \mathbf{y})$.
\begin{algorithm}[!ht]
	\SetAlgoLined
	\textbf{Initialization:}\\
	$s\leftarrow$ socket initialized with port and address\;
	\textit{s.connect}\\
	$\eta, n, N, E \leftarrow s.synchronize()$\\
	$ \{\boldsymbol{w}^{( i)}, \boldsymbol{b}^{( i)}\}_{\forall i\in \{0..l\}} \ \leftarrow \ initialize\ using\ \Phi $\\
	$\{\mathbf{z}^{( i)}\}_{\forall i\in \{0..l\}} ,\{\mathbf{a}^{( i)}\}_{\forall i\in \{0..l\}}\leftarrow \emptyset \ $\\
	$ \left\{\frac{\partial J}{\partial \mathbf{z}^{( i)}}\right\}_{\forall i\in \{0..l\}} ,\left\{\frac{\partial J}{\partial \mathbf{a}^{( i)}}\right\}_{\forall i\in \{0..l\}}\leftarrow \emptyset \ $\\
	\For{$\displaystyle e \ \in \ E $}{
		\For{$\displaystyle \text{each} \ \text{batch}\ ( \mathbf{x},\ \mathbf{y}) \ \text{generated\ from}\ D\ $}{
			$\displaystyle  \mathbf{Forward\ propagation:}$\\
			$\displaystyle \ \ \ \ O.zero\_grad()  $\\
			$\displaystyle \ \ \ \ \mathbf{a}^{0} \ \ \leftarrow \mathbf{x}$ \\
			\For{$i \leftarrow 1$ to $l$}{$\displaystyle \ \ \ \ \mathbf{for} \ i\ \leftarrow \ 1\ \mathbf{to} \ l\ \mathbf{do}$\\
				$\displaystyle \ \ \ \ \ \ \ \ \mathbf{z}^{( i)} \ \leftarrow \ f^{( i)}\left( \mathbf{a}^{( i-1)}\right)$\\
				$\displaystyle \ \ \ \ \ \ \ \ \mathbf{a}^{( i)} \ \leftarrow \ g^{( i)}\left( \mathbf{z}^{( i)}\right)$\\}
			$\displaystyle \ \ \ \  s.send\ ( \mathbf{a}^{(l)})$\\
			$\displaystyle \ \ \ \ s.receive\ ( \mathbf{a}^{(L)})$\\
			$\displaystyle \ \ \ \  \hat{y} \ \leftarrow \ Softmax\left(\mathbf{a}^{( L)}\right)$\\
			$\displaystyle \ \ \ \ J \leftarrow \mathcal{L} (\hat{\mathbf{y}}, \mathbf{y})$\\
			$\displaystyle \mathbf{Backward\ propagation:}$\\
			$\displaystyle \ \ \ \ \text{Compute}\left\{\frac{\partial J}{\partial \hat{\mathbf{y}}}\ \&\ \frac{\partial J}{\partial \mathbf{a}^{(L)}}\right\}$\\
			$\displaystyle \ \ \ \ s.send\ \left( \frac{\partial J}{\partial \mathbf{a}^{(L)}} \right)$\\
			$\displaystyle \ \ \ \ s.receive\ \left( \frac{\partial J}{\partial \mathbf{a}^{( l)}} \right)$\\
			\For{$i\leftarrow 1$ to $l$}{$\text{Compute}\ \left\{ \frac{\partial J}{\partial \boldsymbol{w}^{( i)}}, \ \frac{\partial J}{\partial \boldsymbol{b}^{( i)}} \right\}$\\
				$\displaystyle\ \ \ \ \ \ \ \ \text{Update}\ \boldsymbol{w}^{( i)},\ \boldsymbol{b}^{( i)}$
			}
		}
	}
	\caption{\textbf{Client Side}}
	\label{alg:client}
\end{algorithm}	
The client starts the backward propagation by calculating and sending the gradient of the error w.r.t $\mathbf{a}^{(L)}$, i.e. $\frac{\partial J}{\partial \mathbf{a}^{(L)}}$, to the server. The server continues the backward propagation, calculates $\frac{\partial J}{\partial \mathbf{a}^{(l)}}$ and sends $\frac{\partial J}{\partial \mathbf{a}^{(l)}}$ to the client. After receiving the gradients $\frac{\partial J}{\partial \mathbf{a}^{(l)}}$ from the server, the backward propagation continues to the first hidden layer on the client-side. Note that the exchange of information between client and server in these algorithms takes place in plaintext. As can be seen in~\autoref{alg:client}, the client sends the activation maps $\mathbf{a}^{(l)}$ to the server in plaintext and 
receives the output of the linear layer $\mathbf{a}^{(L)}$ from the server in plaintext. 	
The same applies on the server side: receiving $\mathbf{a}^{(l)}$ and sending $\mathbf{a}^{(L)}$ in the plaintext as can be seen in~\autoref{alg:server}. Sharif \textit{et al.}~\cite{abuadbba2020can} showed that the exchange of plaintext activation maps between client and server using SL 
reveals 
important information regarding the client's raw sequential data. 
Later, in~\autoref{subsec:experiments} 
we show in detail 
how passing the forward activation maps from the client to the server in the plaintext will result in information leakage. 
To mitigate this privacy leakage, we propose the protocol, where the client encrypts the activation maps before sending them to the server, as described in~\autoref{subsubsection: encryptedactivation}.

\begin{algorithm}[h]
	\SetAlgoLined
	\textbf{Initialization:}\\
	$s\leftarrow$ socket initialized with port and address\;
	\textit{s.connect}\\
	$\eta, n, N, E \leftarrow s.synchronize()$\\
	$ \{\boldsymbol{w}^{( i)}, \boldsymbol{b}^{( i)}\}_{\forall i\in \{0..l\}} \ \leftarrow \ initialize\ using\ \Phi $\\
	$\displaystyle \ \ \ \  \{\mathbf{z}^{( i)}\}_{\forall i\in \{l+1..L\}} \leftarrow \emptyset \ $\\
	$\displaystyle \ \ \ \  \left\{\frac{\partial J}{\partial \mathbf{z}^{( i)}}\right\}_{\forall i\in \{l+1..L\}} \leftarrow \emptyset \ $\\
	\For{$\displaystyle e \ \in \ E $}{
		\For{$\displaystyle i \leftarrow 1 \ \mathbf{to} \ N \ $}{
			$\displaystyle \mathbf{Forward\ propagation:}$\\
			$\displaystyle \ \ \ \ O.zero\_grad()  $\\
			$\displaystyle \ \ \ \ s.receive\ (\mathbf{a}^{(l)}) \ \ $ \\
			$\displaystyle \ \ \ \ \mathbf{a}^{(L)} \ \leftarrow \ f^{( i)}\left( \mathbf{a}^{(l)}\right)$\\
			$\displaystyle \ \ \ \ s.send\left( \mathbf{a}^{(L)}\right)$\\
			$\displaystyle \mathbf{Backward\ propagation:}$\\
			$\displaystyle \ \ \ \ s.receive \ \left( \frac{\partial J}{\partial \mathbf{a}^{(L)}}\right)$\\
			$\displaystyle \ \ \ \ \text{Compute}\ \left\{ \frac{\partial J}{\partial \boldsymbol{w}^{(L)}}, \ \frac{\partial J}{\partial \boldsymbol{b}^{(L)}} \right\}$\\
			$\displaystyle \ \ \ \ \text{Update}\ \boldsymbol{w}^{( L)},\ \boldsymbol{b}^{(L)}  $\\
			$\displaystyle \ \ \ \ \text{Compute}\ \frac{\partial J}{\partial \mathbf{a}^{( l)}} $\\
			$\displaystyle \ \ \ \ s.send \left( \frac{\partial J}{\partial \mathbf{a}^{( l)}} \right)$\\
		}
	}
	\caption{\textbf{Server Side}}
	\label{alg:server}
\end{algorithm}

\subsection{Training U-shaped Split 1D CNN with Encrypted Activation Maps}
\label{subsubsection: encryptedactivation}

The protocol for training the U-shaped 1D CNN with a homomorphically encrypted activation map consists of four phases: 
initialization, 
forward propagation, 
classification, 
and backward propagation. The initialization phase 
only takes place 
once at the beginning of the procedure, whereas the other phases 
continue until the model iterates through all epochs.
Each of these phases are described in detail in the following subsections.

\paragraph*{\textbf{Initialization}} The initialization phase consists of socket initialization, context generation, and 
random weight loading. The client first establishes a socket connection to the server and synchronizes the four hyperparameters $\eta ,\ n,\ N, E $ with the server, 
shown in~\autoref{alg:clientHE} and~\autoref{alg:serverHE}. These parameters must be synchronized on both sides to be trained in the same way. Also, the weights on the client and server are initialized with the same set of corresponding weights in the local model to accurately assess and compare the influence of SL on performance. On both the client and the server sides, $\boldsymbol{w}^{(i)}$ are initialized using corresponding parts of $\Phi$. The activation map at layer \textit{i} ($\mathbf{a}^{(i)}$), output tensor of a Conv1D layer ($\mathbf{z}^{(i)}$), and the gradients are initially set to zero. In this phase, 
the context 
generated 
is a specific object that holds encryption keys $\mathsf{pk}$ and $\mathsf{sk}$ of the HE scheme as well as certain additional parameters like $\mathcal{P}$, $\mathcal{C}$ and $\Delta$.

Further information on the HE parameters and how to choose the best-suited parameters can be found in the TenSEAL's benchmarks tutorial\footnote{\url{https://bit.ly/3KY8ByN}}. As 
shown in~\autoref{alg:clientHE} and~\autoref{alg:serverHE}, the context is either public ($\mathsf{ctx_{pub}}$) or private ($\mathsf{ctx_{pri}}$) depending on whether 
it holds the secret key $\mathsf{sk}$. 
Both the $\mathsf{ctx_{pub}}$ and $\mathsf{ctx_{pri}}$ have the same parameters, though 
$\mathsf{ctx_{pri}}$ holds a $\mathsf{sk}$ and 
$\mathsf{ctx_{pub}}$ does not. The server does not have access to the $\mathsf{sk}$ as the client only shares the $\mathsf{ctx_{pub}}$ with the server. After completing the initialization phase, both the client and server proceed to the forward and backward propagation phases.

\paragraph*{\textbf{Forward propagation}} The forward propagation starts on the client side. The client first zeroes out the gradients for the batch of data $(\mathbf{x}, \mathbf{y})$. He then begins calculating the $\mathbf{a}^{(l)}$ activation maps from $\mathbf{x}$, as can be seen in~\autoref{alg:clientHE} where each $f^{(i)}$ is a Conv1D layer.
The Conv1D layer can be described as following: given a 1D input signal that contains $C$ channels, where each channel $\mathbf{x}_{(i)}$ is a 1D array ($i\in \{1,\ldots,C\}$), a Conv1D layer produces an output that contains $C'$ channels. The $j^{th}$ output channel $\mathbf{y}_{(j)}$, where $j\in \{1,\ldots,C'\}$, can be described as\footnote{\url{https://pytorch.org/docs/stable/generated/torch.nn.Conv1d.html}}
\begin{equation}\label{eq:1dconvOp}
	\mathbf{y}_{(j)} = \boldsymbol{b}_{(j)}  + \sum_{i=1}^{C} \boldsymbol{w}_{(i)} \star \mathbf{x}_{(i)},
\end{equation}
where $\boldsymbol{w}_{(i)}, i\in \{1,\ldots,C\}$ are the weights, $\boldsymbol{b}_{(j)}$ are the biases of the Conv1D layer, and $\star$ is the 1D cross-correlation operation. The $\star$ operation can be described as
\begin{equation}
	\mathbf{z}(i) = (\boldsymbol{w} \star \mathbf{x}) (i) = \sum_{j=0}^{m-1}\boldsymbol{w}(j) \cdot \mathbf{x}(i+j), 
\end{equation}
where $\mathbf{z}(i)$ denotes the $i^{th}$ element of the output vector $\mathbf{z}$, and $i$ starts at 0. Here, the size of the 1D weighted kernel is $m$. 

In~\autoref{alg:clientHE}, $g^{(i)}$ can be seen as the combination of Max Pooling and Leaky ReLU functions. The final output activation maps of the $l^{th}$ layer from the client is $\mathbf{a}^{(l)}$. The client then homomorphically encrypts $\mathbf{a}^{(l)}$ and sends the encrypted activation maps $\overbar{\mathbf{a}^{(l)}}$ to the server. In~\autoref{alg:serverHE}, the server receives $\overbar{\mathbf{a}^{(l)}}$ and then performs forward propagation, which is a linear layer evaluated on HE encrypted data $\overbar{\mathbf{a}^{(l)}}$ as
\begin{equation}
	\label{eq:serverHELinear}
	\overbar{\mathbf{a}^{(L)}} = \overbar{\mathbf{a}^{(l)}} \boldsymbol{w}^{(L)} + \boldsymbol{b}^{(L)} .
\end{equation}
After that, the server sends $\overbar{\mathbf{a}^{(L)}}$ to the client~(\autoref{alg:serverHE}). Upon reception, the client decrypts $\overbar{\mathbf{a}^{(L)}}$ to get $\mathbf{a}^{(L)}$, performs Softmax on $\mathbf{a}^{(L)}$ to produce the predicted output $\mathbf{\hat{y}}$ and calculate the loss $J$, as can be seen in~\autoref{alg:clientHE}. 
Having finished the forward propagation 
we may move on to the backward propagation part of the protocol.

\paragraph*{\textbf{Backward propagation}} After calculating the loss $J$, the client starts the backward propagation by initially computing 
$\frac{\partial J}{\partial \hat{\mathbf{y}}}$ and then $\frac{\partial J}{\partial \mathbf{a}^{(L)}}$ and $ \frac{\partial J}{\partial \boldsymbol{w}^{(L)}}$ using the chain rule~(\autoref{alg:clientHE}). 
Specifically, the client calculates
\begin{align}
	\frac{\partial J}{\partial \mathbf{a}^{(L)}} &= \frac{\partial J}{\partial \hat{\mathbf{y}}} \frac{\partial \hat{\mathbf{y}}}{\partial \mathbf{a}^{(L)}}, \text{and} \\
	\frac{\partial J}{\partial \boldsymbol{w}^{(L)}} &= \frac{\partial J}{\partial \mathbf{a}^{(L)}} \frac{\partial \mathbf{a}^{(L)}}{\partial \boldsymbol{w}^{(L)}}.
\end{align}

Following, the client sends $\frac{\partial J}{\partial \mathbf{a}^{(L)}}$ and $ \frac{\partial J}{\partial \boldsymbol{w}^{(L)}}$ to the server. Upon reception, the server computes $\frac{\partial J}{\partial \boldsymbol{b}}$ by simply doing $\frac{\partial J}{\partial \boldsymbol{b}}= \frac{\partial J}{\partial \mathbf{a}^{(L)}}$, based on equation~\eqref{eq:serverHELinear}. The server then updates the weights and biases of his linear layer according to equation~\eqref{equ:serverUpdateWB}.

\begin{algorithm}[!ht]
	\SetAlgoLined
	\textbf{Context Initialization:}\\
	$\displaystyle \ \ \ \ \ \mathsf{ctx_{pri}},\ \leftarrow \ \mathcal{P}, \ \mathcal{C}, \ \Delta, \  \mathsf{pk}, \ \mathsf{sk}$\\
	$\displaystyle \ \ \ \ \ \mathsf{ctx_{pub}},\ \leftarrow \ \mathcal{P}, \ \mathcal{C}, \ \Delta, \  \mathsf{pk}$\\
	$\displaystyle \ \ \ \ \ s.send( \mathsf{ctx_{pub}})$\\
	\For{$\displaystyle e \ \text{in} \ E $}{
		\For{$\displaystyle \text{ each} \ \text{batch}\ ( \mathbf{x},\ \mathbf{y}) \ \text{generated\ from}\ \mathbf{D}\ $}{
			$\displaystyle  \mathbf{Forward\ propagation:}$\\
			$\displaystyle \ \ \ \ O.zero\_grad()  $\\
			$\displaystyle \ \ \ \ \mathbf{a}^{0}\ \ \leftarrow \mathbf{x}$ \\
			\For{$i\ \leftarrow \ 1\ \mathbf{to} \ l$}{
				$\displaystyle \ \ \ \ \ \ \ \ \mathbf{z}^{( i)} \ \leftarrow \ f^{( i)}\left( \mathbf{a}^{( i-1)}\right)$\\
				$\displaystyle \ \ \ \ \ \ \ \ \mathbf{a}^{i} \ \leftarrow \ g^{( i)}\left( \mathbf{z}^{( i)}\right)$}
			$\displaystyle \ \ \ \ \overbar{\mathbf{a}^{(l)}} \ \leftarrow \ \mathsf{HE.Enc}\left(\mathsf{pk}, \mathbf{a}^{(l)}\right)$\\
			$\displaystyle \ \ \ \ s.send \ \overbar{(\mathbf{a}^{(l)})}$\\
			$\displaystyle \ \ \ \  s.receive\ ( \overbar{\mathbf{a}^{(L)})}$\\
			$\displaystyle  \ \ \ \  \mathbf{a}^{( L)} \ \leftarrow \ \mathsf{HE.Dec}\left(\mathsf{sk}, \overbar{\mathbf{a}^{( L)}}\right)$\\
			$\displaystyle  \ \ \ \  \hat{\mathbf{y}} \ \leftarrow \ Softmax\left(\mathbf{a}^{( L)}\right)$\\
			$\displaystyle \ \ \ \  \mathbf{J} \leftarrow \mathcal{L} (\hat{\mathbf{y}}, \mathbf{y})$\\
			$\displaystyle \mathbf{Backward\ propagation:}$\\
			$\displaystyle \ \ \ \ \text{Compute}\left\{\frac{\partial J}{\partial \hat{\mathbf{y}}} \& \frac{\partial J}{\partial \mathbf{a}^{(L)} } \& \frac{\partial J}{\partial \boldsymbol{w}^{(L)}} \right\}$\\
			$\displaystyle \ \ \ \ s.send\left(\frac{\partial J}{\partial \mathbf{a}^{(L)} } \& \frac{\partial J}{\partial \boldsymbol{w}^{(L)}}\right)$\\
			$\displaystyle \ \ \ \ s.receive\left( \frac{\partial J}{\partial \mathbf{a}^{(l)}} \right)$\\
			\For{$i\leftarrow l\ \text{down to} \ 1$}{
				$\displaystyle \ \ \ \ \ \ \ \ \text{Compute} \left\{ \frac{\partial J}{\partial \boldsymbol{w}^{(i)}}, \ \frac{\partial J}{\partial \boldsymbol{b}^{(i)}} \right\}$\\
				$\displaystyle \ \ \ \ \ \ \ \ \text{Update}\ \boldsymbol{w}^{( i)},\ \boldsymbol{b}^{(i)} $}
		}	
	}
	\caption{\textbf{Client Side}}
	\label{alg:clientHE}
\end{algorithm}

\begin{align}
	\label{equ:serverUpdateWB}
	\boldsymbol{w}^{(L)} =  \boldsymbol{w}^{(L)} - \eta\frac{\partial J}{\partial \boldsymbol{w}^{(L)}}, \quad & b^{(L)} = \boldsymbol{b}^{(L)} - \eta\frac{\partial J}{\partial \boldsymbol{b}^{(L)}} \text{.}
\end{align}

\noindent Next, the server calculates
\begin{equation}
	\frac{\partial J}{\partial \mathbf{a}^{(l)}} = \frac{\partial J}{\partial \mathbf{a}^{(L)}} \frac{\partial \mathbf{a}^{(L)}}{\partial \mathbf{a}^{(l)}},
\end{equation}
and sends $\frac{\partial J}{\partial \mathbf{a}^{(l)}}$ to the client. After receiving $\frac{\partial J}{\partial \mathbf{a}^{(l)}}$, the client calculates the gradients of $J$ with respect to the weights and biases of the Conv1D layer using the chain-rule, which can generally be described 
as
\begin{align}
	\label{equ: gradients}
	\frac{\partial J}{\partial \boldsymbol{w}^{(i-1)}} &= \frac{\partial J}{\partial \boldsymbol{w}^{(i)}}\frac{\partial \boldsymbol{w}^{(i)}}{\partial \boldsymbol{w}^{(i-1)}} \\
	\frac{\partial J}{\partial \boldsymbol{b}^{(i-1)}} &= \frac{\partial J}{\partial \boldsymbol{b}^{(i)}}\frac{\partial \boldsymbol{b}^{(i)}}{\partial \boldsymbol{b}^{(i-1)}}    
\end{align}

Finally, after calculating the gradients $\frac{\partial J}{\partial \boldsymbol{w}^{(i)}}, \ \frac{\partial J}{\partial \boldsymbol{b}^{(i)}}$, the client updates $\boldsymbol{w}^{(i)}$ and $\boldsymbol{b}^{(i)}$ using the Adam optimization algorithm~\cite{kingma2014adam}.

\begin{algorithm}
	\SetAlgoLined
	\textbf{Context Initialization:}\\
	$\displaystyle \ \ \ \ \ s.receive(\mathsf{ctx_{pub}})$\\
	\For{$\displaystyle \mathbf{ e} \ \text{in} \ \mathbf{E} $}{
		\For{$\displaystyle i \leftarrow 1 \ \mathbf{to} \ N \ $ }{
			$\displaystyle \mathbf{Forward\ propagation:}$\\
			$\displaystyle \ \ \ \ O.zero\_grad()  $\\
			$\displaystyle \ \ \ \ s.receive\ \overbar{(\mathbf{a}^{(l)})}$\\
			$\displaystyle \ \ \ \ \overbar{\mathbf{a}^{(L)}} \ \leftarrow \ \mathsf{HE.Eval} \left(f^{( i)}\left( \overbar{\mathbf{a}^{( l)}}\right)\right)$\\
			$\displaystyle \ \ \ \ s.send\left( \overbar{\mathbf{a}^{( L)}}\right)$\\
			$\displaystyle \mathbf{Backward\ propagation:}$\\
			$\displaystyle \ \ \ \ s.receive\left\{\frac{\partial J}{\partial \mathbf{a}^{(L)}} \& \frac{\partial J}{\partial \boldsymbol{w}^{(L)}}\right\}$\\
			$\displaystyle \ \ \ \ \text{Compute}\ \frac{\partial J}{\partial \boldsymbol{b}^{(L)}} $\\
			$\displaystyle \ \ \ \  \text{Update}\ \boldsymbol{w}^{(L)},\ \boldsymbol{b}^{(L)} $\\
			$\displaystyle \ \ \ \ \text{Compute}\ \frac{\partial J}{\partial \mathbf{a}^{(l)}} $\\
			$\displaystyle \ \ \ \ s.send \left( \frac{\partial J}{\partial \mathbf{a}^{(l)}} \right)$\\
		}
	}
	\caption{\textbf{Server Side}}
	\label{alg:serverHE}
\end{algorithm}

Note that in the backward pass, by sending both $\frac{\partial J}{\partial \mathbf{a}^{(L)}}$ and $\frac{\partial J}{\partial \boldsymbol{w}^{(L)}}$ to the server, we help the server keep his parameters in plaintext and prevent the multiplicative depth of the HE from growing out of bound, however, this leads to a privacy leakage of the activation maps.

\section{Performance Analysis}
\label{sec:performance}

We evaluate our method on 
the MIT-BIH dataset~\cite{moody2001impact}. 

\begin{figure}[!h]
	\centering
	\includegraphics[width=0.48\textwidth]{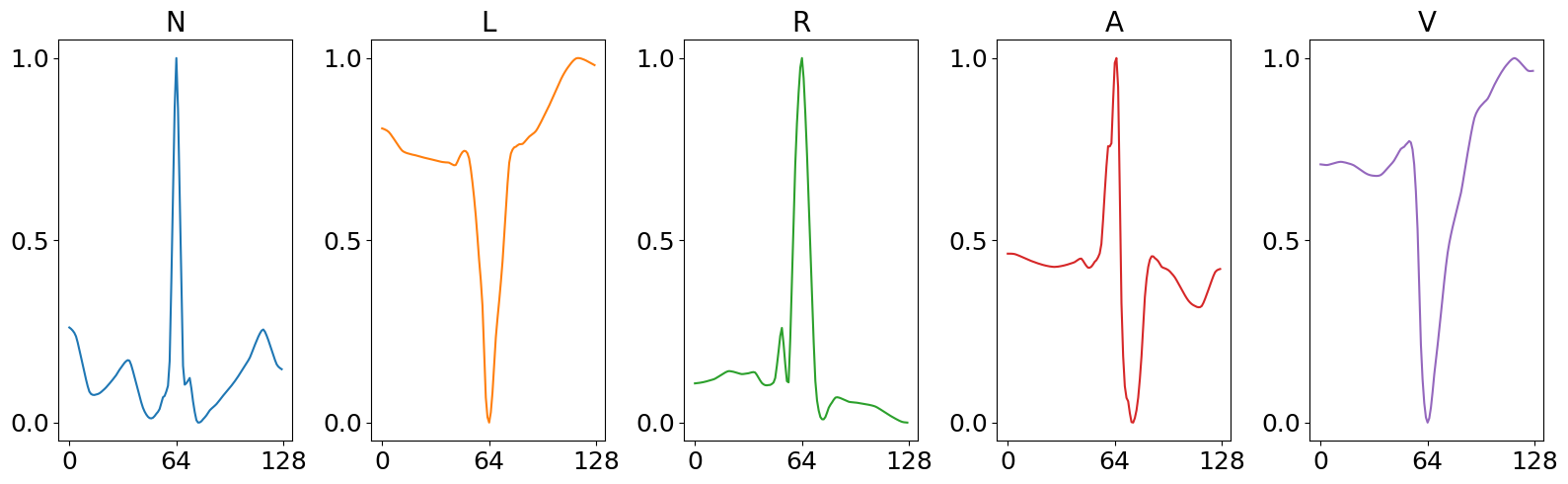}
	\caption{Heartbeats from the processed ECG dataset.}
	\label{fig:ECGDataset}
\end{figure}

\smallskip
\noindent \textbf{MIT-BIH}
We use the pre-processed dataset from~\cite{abuadbba2020can}, which is based on the MIT-BIH arrhythmia (abnormal heart rhythm) database~\cite{moody2001impact}. The processed dataset contains 26,490 samples of heartbeat that belong to 5 different types: N (normal beat), L (left bundle branch block), R (right bundle branch block), A (atrial premature contraction), V (ventricular premature contraction). An example heartbeat 
of each class is visualized in \autoref{fig:ECGDataset}.

To train our network, the dataset is then split into a train and test split according to~\cite{abuadbba2020can}. This results in both the train and test split as matrices of size $[13245, 1, 128]$, meaning that they contain 13,245 ECG samples, each sample has one channel and 128 timesteps.

\paragraph*{\textbf{Experimental Setup}}
\label{subsec:setup}
All neural networks are trained on a machine with Ubuntu 20.04 LTS, processor Intel Core i7-8700 CPU at 3.20GHz, 32Gb RAM, GPU GeForce GTX 1070 Ti with 8Gb of memory. We write our program in the~\href{https://www.python.org/downloads/release/python-397/}{Python programming language version  3.9.7}. The neural nets are constructed using the~\href{https://pytorch.org/get-started/previous-versions/}{PyTorch library version 1.8.1+cu102}. For HE algorithms, we employ the~\href{https://github.com/OpenMined/TenSEAL}{TenSeal library version 0.3.10}. We perform our experiments in the localhost setting. 

In terms of hyperparameters, we train all networks with 10 epochs, $\eta=0.001$ learning rate, and $n=4$ training batch size. 
For the split neural network with HE activation maps, we use the Adam optimizer for the client model and mini-batch Gradient Descent for the server. 
We use GPU for 
networks 
trained on the plaintext. For the U-shaped SL model on HE activation maps, we train the client model on GPU, and the server model on CPU.

\subsection{Evaluation}
\label{subsec:experiments}
In this section, we report the experimental results in terms of accuracy, training duration and communication throughput. We measure the accuracy of the neural nets on the plaintext test set after the training processes are completed. 
The 1D CNN models used on MIT-BIH dataset have two Conv1D layers and one linear layer. The activation maps are the output of the last Conv1D layer. 

We experiment with the activation maps  of $[\text{batch size}, 256]$ for the MIT-BIH dataset. 
We denote the 1D CNN model with an activation map sized $[\text{batch size}, 256]$ as $M_1$.

\begin{figure}[!hb]
	\centering
	\includegraphics[width=0.48\textwidth]{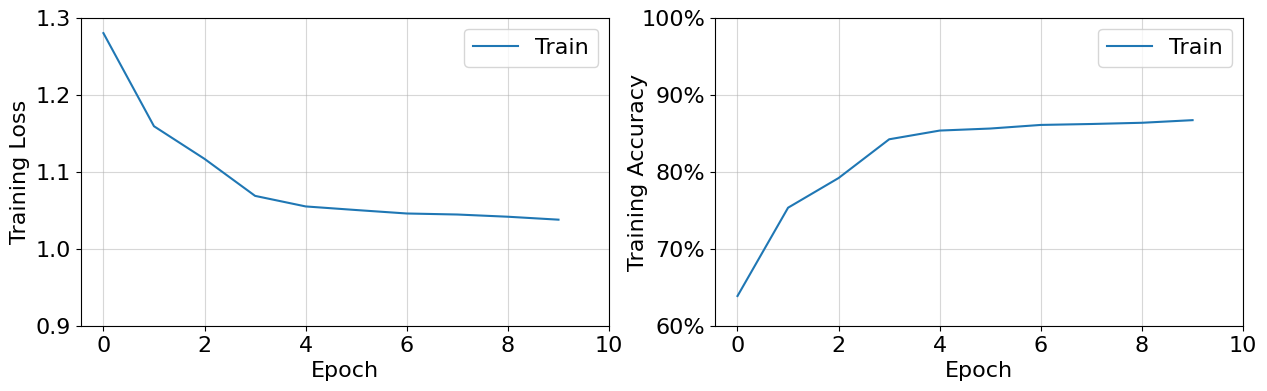}
	\caption{Results when training locally on the plaintext MIT-BIH dataset with activation maps of size $[\text{batch size}, 256]$.}
	\label{fig:localTrain256}
\end{figure}

\smallskip
\noindent \textbf{Training Locally} 
Results when training $M_1$ locally on the MIT-BIH plaintext dataset are shown in \autoref{fig:localTrain256}. The neural network 
learns quickly and is able to decrease the loss drastically from epoch 1 to 5. 
From epoch 6-10, the loss begins to plateau. After training for 10 epochs, we test the trained neural network on the test dataset and get 88.06\% accuracy. Training the model locally on plaintext takes 4.8sec for each epoch on average.

\smallskip
\noindent \textbf{U-shaped Split Learning using Plaintext Activation Maps}
Our experiments, 
show that training the U-shaped split model on plaintext (reported in section~\ref{subsec:u_shaped_split}) 
produces the same results in terms of accuracy compared to local training 
for model $M_1$. This result is similar to the findings of 
~\cite{abuadbba2020can}. Even though the authors of~\cite{abuadbba2020can} only used the vanilla version of the split model, they too found that, compared to training locally, 
accuracy was not reduced.

We will now discuss the training time and communication overhead of the U-shaped split models and compare them to their local versions. For the split version of $M_1$, each training epoch takes 8.56 seconds on average, 
hence 43.9\% longer than local training. 
The U-shaped split models take longer to train due to the communication between the client and the server. The communication cost for one epoch of training split $M_1$ 
is 33.06 Mb.

\begin{figure}[!hb]
	\centering
	\includegraphics[width=0.48\textwidth]{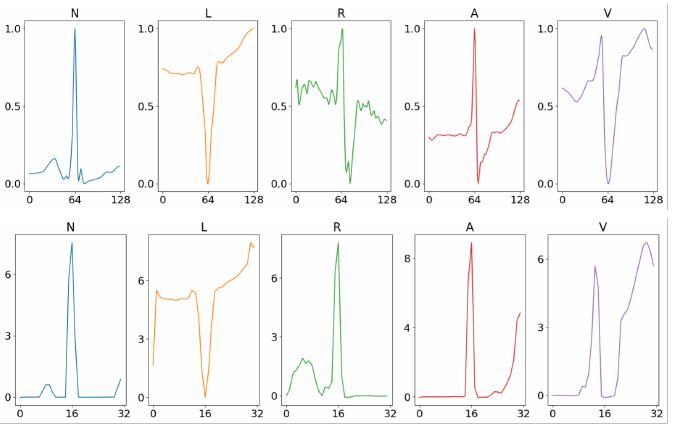}
	\caption{Top: client input data. Bottom: one of the output channels from the $M_1$ model's second convolution layer.} 
\label{fig:visual_invertibility}
\end{figure}

\smallskip
\noindent \textbf{Visual Invertibility}
\label{subsec:VisualInvert}
In the SL model, the activation maps are sent from 
client to 
server to continue the training process. 
A visual representation of the activation maps reveals 
a high similarity between certain activation maps and the input data from the client, as demonstrated in \autoref{fig:visual_invertibility} for the models trained on the MIT-BIH dataset. 
The figure indicates that, compared to the raw input data from the client (the first row of~\autoref{fig:visual_invertibility}), some activation maps (as plotted in the second row of~\autoref{fig:visual_invertibility}) have exceedingly similar patterns. This phenomenon clearly compromises the privacy of the client's raw data. The authors of~\cite{abuadbba2020can} quantify the privacy leakage by measuring the correlations between the activation maps and the raw input signal by using two metrics: distance correlation and Dynamic Time Warping. 
This approach allows them to measure whether their solutions 
mitigate privacy leakage work. Since our work uses HE, 
said metrics are unnecessary as  
the activation maps are encrypted.

\smallskip
\noindent \textbf{U-shaped Split 1D CNN with Homomorphic Encrypted Activation Maps}
\label{subsec:UShapedHE}
We train the split neural networks $M_1$ on the MIT-BIH dataset using encrypted activation maps according to \autoref{subsubsection: encryptedactivation}. To encrypt the activation maps on client side (i.e.\ before sending them to the server), we experiment with five different sets of HE parameters for model $M_1$. 
Additionally, we perform experiments using different combinations of HE parameters. \autoref{tab:trainingTestingResultsMITBIH} shows the results in terms of training time, testing accuracy, and communication overhead for the neural networks with different configurations. 
For the U-shaped SL version on the plaintext, we captured all communication between client and server. For training split models on encrypted activation maps, we approximate the communication overhead for one training epoch by getting the average communication of training on the first ten batches of data, then multiply that with the total number of training batches.

For the $M_1$ model, 
the best test accuracy was 85.41\%, when using the HE parameters with polynomial modulus $\mathcal{P}=4096$, coefficient modulus $\mathcal{C}=[40, 20, 20]$, scale $\Delta=2^{21}$. 
The accuracy drop was $2.65\%$ compared to training the same network on plaintext. 
This set of parameters achieves higher accuracy compared to the bigger sets of parameters with $\mathcal{P}=8192$, while requiring much lower training time and communication overhead. The result when using the first set of parameters with $\mathcal{P}=8192$ is close ($85.31\%$), but with a much 
longer training time (3.67 times longer) and communication overhead (8.43 times 
higher). 

\begin{table*}[ht]
\centering
\caption{Training and testing results on the MIT-BIH dataset. Training duration and communication are reported per epoch.}
\resizebox{\textwidth}{!}{%
	\begin{tabular}{c|c|cccc|c|c|c}
		\hline
		\multirow{2}{*}{Network} & \multirow{2}{*}{Type of Network} & \multicolumn{4}{c|}{HE Parameters}                                                                                      & \multirow{2}{*}{Training duration (s)} & \multirow{2}{*}{Test accuracy (\%)} & \multirow{2}{*}{Communication (Tb)} \\ \cline{3-6}
		&                                  & \multicolumn{1}{c|}{BE}                     & \multicolumn{1}{c|}{$\mathcal{P}$}    & \multicolumn{1}{c|}{$\mathcal{C}$}             & $\Delta$        &                                    &                                     &                                     \\ \hline
		\multirow{9}{*}{$M_1$}     & Local                            & \multicolumn{4}{c|}{}                                                                                                   & 4.80                                   & 88.06                               & 0                                   \\ \cline{2-9} 
		& Split (plaintext)                & \multicolumn{4}{c|}{}                                                                                                   & 8.56                                   & 88.06                               & 33.06e-6                            \\ \cline{2-9} 
		& \multirow{5}{*}{Split (HE)}     & \multicolumn{1}{c|}{\multirow{5}{*}{False}} & \multicolumn{1}{c|}{8192} & \multicolumn{1}{c|}{[60,40,40,60]} & $2^{40}$ & 50 318                                 & 85.31                               & 37.84                               \\ \cline{4-9} 
		&                                  & \multicolumn{1}{c|}{}                       & \multicolumn{1}{c|}{8192} & \multicolumn{1}{c|}{[40,21,21,40]} & $2^{21}$ & 48 946                                 & 80.63                               & 22.42                               \\ \cline{4-9} 
		&                                  & \multicolumn{1}{c|}{}                       & \multicolumn{1}{c|}{4096} & \multicolumn{1}{c|}{[40,20,20]}    & $2^{21}$ & 14 946                                 & 85.41                               & 4.49                                \\ \cline{4-9} 
		&                                  & \multicolumn{1}{c|}{}                       & \multicolumn{1}{c|}{4096} & \multicolumn{1}{c|}{[40,20,40]}    & $2^{20}$ & 18 129                                 & 80.78                               & 4.57                                \\ \cline{4-9} 
		&                                  & \multicolumn{1}{c|}{}                       & \multicolumn{1}{c|}{2048} & \multicolumn{1}{c|}{[18,18,18]}    & $2^{16}$ & 5 018                                  & 22.65                               & 0.58                                \\ \hline
	\end{tabular}%
}
\label{tab:trainingTestingResultsMITBIH}
\end{table*}

Uur experiments show that training on encrypted activation maps can produce
optimistic results, with accuracy dropping by 2-3\% for the best sets of HE parameters. 

The set of parameters with $\mathcal{P}=8192$ achieve the second highest test accuracy, though incurring the highest communication overhead and the longest training time. The set of parameters with $\mathcal{P}=4096$ can offer a good trade-off as they can produce on-par accuracy with $\mathcal{P}=8192$, while requiring 
significantly less communication and training time. Experimental results show that with the smallest set of HE parameters $\mathcal{P}=2048$, $\mathcal{C}=[18, 18, 18]$, $\Delta=2^{16}$, 
the least amount of communication and training time is required.

\section{Conclusion}
\label{sec:Conclusion}

This paper focused on how to train ML models in a privacy-preserving way using a combination of split learning -a promising machine-learning method- and homomorphic encryption. We  constructed protocols by which a client and a server could collaboratively train a model without revealing significant information about the raw data. As far as we are aware, this is the first time split learning is used on encrypted data. 

\bibliographystyle{abbrv}
\bibliography{split-ways}
\end{document}